\documentclass[reprint,prd,amsmath,twocolumn,amssymb,aps,bibliography=totoc]{revtex4-1}

\usepackage{dcolumn}
\usepackage{bm}
\usepackage{notes2bib}
\usepackage{appendix}
\usepackage{amsmath}
\usepackage{slashed}
\usepackage{multirow}
\usepackage{color}
\usepackage{mathtools}
\usepackage{graphicx}
\usepackage{braket}
\usepackage{setspace}
\usepackage{graphicx}
\usepackage{geometry}
\usepackage{hyperref}
\usepackage{starfont}
\usepackage{amssymb}
\usepackage{bbold}
\usepackage[shortlabels]{enumitem}
\usepackage{framed}
\usepackage{empheq}
\usepackage{amsmath}
\usepackage{tikz-feynman}
\usepackage{simpler-wick}
\usepackage{wasysym}
\usepackage{mhchem}

\def\Mp{M_{\rm Pl}}
\def\Mpl{M_{\rm Pl}}

\def\TeV{\,{\rm TeV}}
\def\GeV{\,{\rm GeV}}
\def\MeV{\,{\rm MeV}}

\def\eV{\,{\rm eV}}
\def\Tos{T_{\rm os}}

\def\Tra{T_{\rm ra}}

\def\tpp{t_{\rm pp}}

\def\Hpp{H_{\rm pp}}
\def\Hra{H_{\rm ra}}

\def\mub{\Lambda_{\rm br}}
\def\rx{r_\xi}
\def\thX{\dot \theta_X}
\DeclarePairedDelimiter\abs{\lvert}{\rvert}
\makeatletter
\let\oldabs\abs
\def\abs{\@ifstar{\oldabs}{\oldabs*}}
\makeatother

\newcommand{\bea}{\begin{eqnarray}}
\newcommand{\eea}{\end{eqnarray}}

\begin{document}
\title{Coherent relaxion dark matter}

\author{Abhishek Banerjee}
\affiliation{Department of Particle Physics and Astrophysics,\\
Weizmann Institute of Science, Rehovot 7610001, Israel}

\author{Hyungjin Kim}
\affiliation{Department of Particle Physics and Astrophysics,\\
Weizmann Institute of Science, Rehovot 7610001, Israel}

\author{Gilad Perez}
\affiliation{Department of Particle Physics and Astrophysics,\\
Weizmann Institute of Science, Rehovot 7610001, Israel}

\begin{abstract}
We show that relaxion, that addresses the hierarchy problem, can account for the observed dark matter (DM) relic density.
The setup is similar to the case of axion DM models topped with a dynamical misalignment mechanism.
After the reheating, when the temperature is well above the electroweak scale, the backreaction potential disappears and the relaxion is displaced from its vacuum. When the ``wiggles" reappear the relaxion coherently oscillates around its minimum as in the case of vanilla axion DM models. We identify the parameter space such that the relaxion is retrapped leading to the standard cosmology.
When the relaxion is lighter than $10^{-7}\,$eV, Hubble friction during radiation-domination is sufficiently strong for retrapping, and even minimal models are found to be viable. It also leads to a new constraint on relaxion models, as a certain region of their parameter space could lead to overabundant relaxion DM. 
Alternatively, even a larger parameter space exists when additional friction is obtained by particle production from additional coupling to an additional dark photon field.
The phenomenology of this class of models is quite unique, as it implies that we are surrounded by a time-dependent axion-like field that due to relaxion-Higgs mixing implies time-dependent Higgs vacuum-expectation-value that lead to time-variation of all coupling constants of nature. 
\end{abstract}
\maketitle

{\bf Introduction.}
The relaxion mechanism provides an alternative solution to the Higgs naturalness problem~\cite{Graham:2015cka}.
Within the relaxion framework, the electroweak (EW) scale is not a fundamental scale of a UV theory, but emerges as a result of dynamical evolution of our universe.
The Higgs mass is not a constant but rather a time dependent function of an axion-like field, the relaxion.
An example of the potential of relaxion and Higgs that realizes the relaxion mechanism is~\bibnote{We focus here on Hubble-friction based models. See~\cite{Hook:2016mqo,Fonseca:2018xzp} for alternatives.} 
\bea
\!\! \! \!\! \! \! \!\!  V(H,\phi) \!= (\Lambda^2 - g\Lambda \phi) |H|^2 - c g \Lambda^3 \phi - \mu^2 |H|^{2} \cos{\phi\over f}\,,
\label{rel_potential}
\eea
where $\Lambda$ is the cutoff scale for the Higgs mass, $f$ is axion decay constant, $c$ is an order one coefficient, $g\sim \mu^2 v^2/f\Lambda^3$, and $\mu$ is the scale characterizing the backreaction potential, with $v$ being the EW scale. 
In this scenario, the electroweak scale is scanned during the inflationary phase of the universe.
Initially, the Higgs mass is positive and thus the vacuum expectation value (VEV) of the Higgs vanishes. 
As the relaxion evolves along its potential, the Higgs mass decreases, and at some point, it becomes negative from which nonzero VEV is generated.
Once the Higgs developes nonvanishing VEV, the backreaction potential, the last term in Eq.~\eqref{rel_potential}, appears to feedback the evolution of the relaxion.
When the Higgs mass approaches the electroweak scale, the backreaction potential balances the relaxion potential, and thus, stops the relaxion from further evolution.
This close interplay between the relaxion rolling potential and the Higgs-dependent backreaction potential allows the relaxion to be stabilized at the vacuum that provides electroweak scale with UV parameters chosen in a technically natural way (for explicit realizations and further discussions, see for instance~\cite{Graham:2015cka,Espinosa:2015eda, Gupta:2015uea, Batell:2015fma,Choi:2015fiu,Kaplan:2015fuy,Davidi:2018sii}).

In this paper, we investigate whether the model presented above can account for the observed dark matter relic density, in the context of the standard $\Lambda$CDM cosmology with a reheating temperature well above the electroweak scale. 
The only non-SM light degree of freedom in a minimal scenario is the relaxion field itself. It is shown that via a dynamical misalignment mechanism, the relaxion follows a viable axion-like DM evolution. 
 \\
 
{\bf Basic idea.}
The idea is based on the following observation.
During inflation, the relaxion scans the electroweak Higgs mass, and settles down at one of its local minima.
If the universe is reheated with temperature above the critical temperautre of EW phase transition, the EW symmetry is restored, and the backreaction potential disappears.
As a result, the relaxion field begins to evolve again, until the backreaction potential appears at some temperature, $\Tra$.
Requiring that the relaxion has been trapped in a close-by minima, the relaxion field is displaced from its local mininum with a certain misalignment angle, $\Delta\theta \!= \! \Delta \phi /f$.
Consequently, when the Hubble scale drops below its mass, it begins to oscillate around the minimum. 
This coherently oscillating relaxion field eventually constitutes the DM in the present universe.
To guarantee such a ``relaxion-miracle" can occur, we must ensure that the relaxion is trapped again by the backreaction potential after its second evolution during the radiation-dominated universe.
We show below that a light relaxion can be efficiently trapped either via the Hubble friction during the radiation domination era in a truly minimal model, or via particle production from relaxion coupling to dark photon~\bibnote{A possibility of the relaxion DM is briefly mentioned in the original paper~\cite{Graham:2015cka} without any explicit computations. Although it is possible that the relaxion could obtain nonvanishing misalignment angle after the reheating in their QCD model~\cite{Graham:2015cka} since the linear potential also couples to the inflaton field, it is different from our consideration as the dynamical misalignment mechanism that we are considering does not invoke any needs of coupling the rolling potential to the inflaton field.}. \\

{\bf Minimal model.}
A viable DM model would require that the relaxion is sufficiently misaligned from local minimum to reproduce the observed relic abundance, and that the backreaction potential should be able to trap the relaxion while maintaining the Higgs mass close to its original value set by the dynamics during inflation.

To quantify the required size of misalignment angle, we consider the dark matter density at the beginning of oscillation for a given misalignment angle, $m_\phi Y_\phi(t_{\rm os}) = \rho_\phi(t_{\rm os})/s(t_{\rm os}) \! \simeq \! m_\phi^2 f^2 (\Delta\theta)^2/2 s(t_{\rm os})$, where $t_{\rm os}$ is the time at which the oscillation begins, and $s(t)$ is the entropy density.
The temperature of the universe when the relaxion starts to oscillate is given as
$\Tos \sim \min \left[\Tra, \,  \sqrt{m_\phi \Mp}\right]$, where $\Tra$ is the temperature of the universe when the backreaction potential appears again, and the second term in the squared parenthesis is obtained by $3H(T_{\rm os}) = m_\phi$.
The resulting relic abundance at the present universe is $\Omega_\phi = m_\phi Y_\phi s(t_0) / \rho_{\rm crit}$, which can be written as 
\bea
\!\!\!\Omega_\phi h^2 &\approx& 3 \times  (\Delta \theta)^2_{T=\Tos} \left( \frac{\sqrt{m_{\phi}f}}{1\GeV} \right)^4 \left( \frac{100\GeV}{\Tos} \right)^3,
\label{relic}
\eea
where the observed DM abundance is $\Omega_{\rm DM}h^2 \simeq 0.12$~\cite{Aghanim:2018eyx}.
We assume that the relaxion is trapped right after the reappearance of the backreaction potential, i.e. at $T=T_{\rm ra}$.

To realize nonvanishing misalignment angle, we consider the universe with a reheating temperature well above the critical temperature of EW phase transition.
In this case, as we have already portrayed above, the relaxion begins to evolve after the reheating since the Higgs-dependent backreaction potential has disappeared.
The evolution is governed by the relaxion equation of motion,
$\ddot{\phi} + 3 H \dot{\phi} - g \Lambda^3 \simeq 0$, 
where the solution is $\dot{\phi}(t) = \frac{2}{5} g \Lambda^3 t [ 1 - (t_{\rm rh}/t)^{5/2}]$ with a proper time at the reheating $t_{\rm rh}$. 
Because of this evolution, the relaxion can naturally be misaligned from its local minimum by $\Delta\phi(t)/f = m_0^2 / 20 H^2(t)$ modulo $2\pi$, where $m_0^2 = \mu^2 v^2 / f^2 \equiv \Lambda_{\rm br}^4/f^2$.
Note that this mass scale $m_0^2$ is different from the physical mass of relaxion $m_\phi^2$ at each local minimum.
We discuss this in more detail below.

Another necessary condition for relaxion DM is that the backreaction potential should be able to trap the relaxion, once it appears again at some temperature $T=T_{\rm ra}$~\bibnote{Throughout this paper, we assume that the backreaction potential appears right at the barrier reappearance temperature, $T = \Tra$. This might not be strictly true whenever the backreaction potential arises from a confinement of QCD-like hidden gauge force. Especially in the case of QCD, topological susceptibility has nontrivial temperature dependence, and so does the mass of axion. We ignore this nontrivial temperature dependence for an order of magnitude estimation on relaxion dark matter relic abundance.}. 
Finding a relaxion-trapping condition is not possible without describing the detail shape of the relaxion potential around its local minimum because it crucially depends on the potential height and the separation between local extremum. 
For this reason, we shall briefly discuss the detail shape of the relaxion potential around its local minima~\bibnote{We thank the anonymous referee for raising this issue to us.}.

To determine the relevant part of the relaxion potential, we trace back to the relaxion evolution during the inflation.
In the original relaxion scenario, the relaxion stops its classical evolution when the slope of the backreaction potential, $V'_{\rm br}(\phi) \! = \! - (\mu^2 v^2 / f) \sin(\phi/f)$, becomes larger than the slope of the rolling potential, $V'_{\rm roll}(\phi) = - g \Lambda^3$, for the first time.
The potential height $\Delta V$ and the separation $(\Delta \theta)_{\rm sep}$ between local extremum at the stopping point are particularly important for our discussion, as they determine the maximum misalignment angle and the maximum energy density of the relaxion field.
When the classical stopping condition, $V'_{\rm roll} + V'_{\rm br} = 0$, is met for the first time, we find
\bea
\Delta V = \delta^3 \Lambda_{\rm br}^4,
\qquad
(\Delta \theta)_{\rm sep} = 2 \delta,
\qquad
m_\phi^2 = \delta m_0^2,
\label{rel_params}
\eea
where $\delta = \mu / \Lambda$, $m_\phi$ is the physical relaxion mass, and $\Delta V$ and $(\Delta \theta)_{\rm sep}$ are the potential height and the distance between the first local minimum and maximum, respectively.
Note that all quantities are suppressed by additional small parameter $\delta$ compared to what we would naively expect from $V_{\rm br}$.
This is due to the fine scanning of electroweak scale; the Higgs VEV changes by $\Delta v^2 / v^2 \simeq \delta^2$ for each $\Delta \phi \sim f$, and thus, the slope of backreaction potential increases only incrementally.
Thus, the first two solutions of $V' = V'_{\rm roll} + V'_{\rm br} =0$ are close to each other in the field space by $\delta$, and the mass at the first local minimum is also suppressed by some power of $\delta$ because the distance between local minimum and the inflection point, which is a solution of $V''=0$, vanishes as $\delta \to 0$. 
Similar argument also applies for $n$-th local minimum by replacing $\delta \to \delta_n = \sqrt{n}\delta$.
We provide more detailed arguments on this in Appendix.~\ref{sec:app2}.

The potential height, the physical relaxion mass, the separation between local extremum  depend on the minimum at which the relaxion is stabilized, which are all relevant for the discussion of relaxion dark matter.
Thus, it is crucial to correctly identify the minimum that the relaxion is stabilized by the end of the inflation, which will be the initial condition of relaxion evolution after the reheating.
If the Hubble expansion parameter during inflation is smaller than the potential height at the first minimum, $\Delta V > H_I^4$, the tunneling rate is exponentially suppressed, so that the relaxion is likely to be stabilized at the first minimum.
In the other case, $\Delta V < H_I^4$, the relaxion tunnels to the point where $\Delta V \simeq H_I^4$ due to the quantum tunneling.
As we will see below, for the parameter space where the relaxion DM can be realized, the potential height at the first minimum is larger than $H_I^4$, even if we take the largest inflationary Hubble parameter that is allowed by relaxion scenario.
By this reason, we will assume below that the relaxion begins to evolve from the first local minimum after reheating.

Now, we discuss the relaxion-trapping condition.
When the backreaction potential appears again after the reheating, the slope of the total relaxion potential would be $\sim \delta^2 \Lambda_{\rm br}^4/f$ from the local minimum to the local maximum, while the relaxion velocity at this moment is $H\dot{\phi} \sim \Lambda_{\rm br}^4 /f \gg \delta^2 \Lambda_{\rm br}^4 /f$.
Thus, for the evolution from the local minimum to the maximum, we may ignore the potential slope, and, from the equation of motion, $\ddot{\phi} + 3 H \dot{\phi} \simeq 0$, the relaxion velocity quickly redshifts as $\dot{\phi} \propto a^{-3}$.
The subsequent evolution of relaxion is dominantly determined by its velocity at $T_{\rm ra}$, and if the total field excursion of relaxion is smaller than $(\Delta \theta)_{\rm sep}$, the relaxion will be trapped by the backreaction potential.
The total field excursion is
\bea
(\Delta\theta)_{\rm mis} \simeq \int_{t_{\rm rh}}^{\infty} dt \frac{\dot{\phi}}{f} = \frac{1}{4} \frac{m_0^2}{H_{\rm ra}^2}.
\label{mis}
\eea
Thus, we impose the relaxion stopping condition as
\bea
(\Delta \theta)_{\rm mis} < (\Delta \theta)_{\rm sep} = 2 \delta.
\label{trapping}
\eea
Note that $(\Delta \theta)_{\rm mis}$ is the misalignment angle that determines the relic abundance in Eq.~\eqref{relic}. 
Given $(\Delta \theta)_{\rm mis} < (\Delta \theta)_{\rm sep}$, the relaxion evolves less than $\Delta \phi = 2\pi f$, and thus the Higgs mass does not change after reheating in this minimal scenario.

From the relaxion-trapping condition, Eq.~\eqref{trapping}, the mass of physical relaxion is always smaller than the Hubble expansion parameter at $T= T_{\rm ra}$.
This indicates that the coherent oscillation begins after the reappearance of the backreaction potential, i.e., $T_{\rm os} \sim (m_\phi \Mp)^{1/2} < T_{\rm ra}$. 
As a consequence, the relic abundance depends on relaxion parameters as $\Omega_\phi h^2 \propto m_\phi^{19/6} f^{2/3}$, which has a sharp dependence on the relaxion mass. 
We could also find a relaxion DM window for the minimal scenario as
\bea
5\times10^{-11} \, \Lambda_{\rm TeV}^{\frac{8}{11}} \Big(\frac{\Mp}{f}\Big)^{\frac{20}{11}}
\lesssim \frac{m_\phi}{\rm eV} \lesssim 
10^{-8} \, \Lambda_{\rm TeV}^{-\frac{4}{7}} \Big( \frac{T_{\rm ra}}{150\GeV} \Big)^{\frac{20}{7}} \!\!,
\nonumber\\
\eea
where $\Lambda_{\rm TeV} =\Lambda/$TeV. 
For the upper bound, we take $f$ such that the misalignment angle saturates to its maximum value, while for the lower bound, we require $f$ to be sub-Planckian and choose the smallest $T_{\rm ra}$ such that we still satisfy the relaxion-trapping condition.
When $\Lambda \simeq 1\TeV$ and $f=\Mp$, the smallest relaxion DM mass is around $5\times10^{-11}\eV$, and $T_{\rm ra} \simeq 10 \GeV$.

We finish this section by noting that there is a certain region that potentially lead to a relaxion overabundance, resulting in a new constraint on minimal models:
\bea
\!\!\!\!\!\! m_\phi \lesssim10^{-8} \eV \left( {\Tra\over v} \right)^{48\over 19} \left( \frac{\Mpl}{f} \right)^{\frac{4}{19}}\left( {10 \TeV\over \Lambda } \right)^{\frac{8}{19}} \label{ovab}
\!\!\!.
\eea

{\bf Dissipation from dark photons.}
The main difficulty for heavier relaxion is that the field velocity $\dot{\phi}$ is too large to be trapped by the backreaction potential. 
However, the model's parameter space can be extended to heavier relaxion mass once we introduce couplings to new fields, which are responsible for additional dissipational channel for relaxion.
In particular, adding a coupling to dark photons via the operator $ (r_X/4f) \phi X_{\mu\nu} \widetilde{X}^{\mu\nu}$, with the dark photon field strength $X_{\mu\nu} = \partial_\mu X_\nu - \partial_\nu X_\mu$, and its dual $\widetilde{X}$, would lead to a new source of dissipation (see {\it e.g.}~\cite{Choi:2016kke,Tangarife:2017vnd}).
The new interaction introduces a source term to the equation of motion for relaxion,
\bea
\ddot{\phi} + 3 H \dot{\phi} + \frac{\partial V(v,\phi)}{\partial \phi} = - \frac{r_X}{4f a^4} \langle X_{\mu\nu} \widetilde{X}^{\mu\nu} \rangle,
\label{eom_dp}
\eea
providing additional channel for the relaxion to dissipate its kinetic energy.
At the same time, nonzero kinetic energy triggers an exponential production of dark photons.
To illustrate this point, we first expand the dark photon field in Fourier space,
$$
\vec{X}(\tau,\vec{x}) = \int \frac{d^3k}{(2\pi)^3} \sum_{\lambda = \pm} \left[ \vec{\epsilon}_\lambda(\hat{k}) a_{\vec{k},\lambda} e^{i\vec{k}\cdot\vec{x}} X_\lambda(\tau,\vec{k}) + {\rm h.c.} \right],
$$
and find the equation of motion for dark photon as
$X_\pm'' + (k^2 \mp  r_X k \theta' ) X_\pm = 0\,,$
where the prime denotes a derivative with respect to the conformal time, $d\tau = dt /a(t)$, and $\theta\equiv \phi/f$. 
The nonvanishing classical background, $\theta'\neq0$, leads to exponential production of one of the helicity modes.
Thus, eventually, the source term in the equation of motion becomes comparable to the other terms, effectively alleviating the slope of relaxion potential. 
As a result, the relaxion field-velocity approaches an asymptotic value~\cite{Choi:2016kke}, $|\thX| \equiv \rx H,$
where the coefficient $\rx$ depends on relaxion parameter only logarithmically, and numerically takes a value around $\xi\equiv\rx\, r_X \sim {\cal O}(10)$~\bibnote{For our purpose, we ignore the possible logarithmic time dependence of $\dot{\phi}$. 
We provide more detailed arguments on asymptotic behavior of relaxion evolution in Appendix~\ref{sec:app} together with an estimation on logarithmic time dependence of $\dot{\phi}$.}.
The time scale that this particle production kicks in is $2\tpp= \Hpp^{-1} \simeq \left( 5 \rx \right)^{1/2} m_0^{-1}$.
Once particle production becomes efficient, the relaxion kinetic energy is a decreasing function in time.

In the presence of dark photon, we find the total field excursion from the reheating to the temperature at which the backreaction potential appears again as
\bea
\!\!\!\!\!\!\!\!\! \frac{\Delta \phi}{f} \simeq \frac{m_0^2}{20 H_{\rm pp}^2} + \frac{\rx}{2} \ln\left(\Hpp\over \Hra\right) \!
\simeq \frac{\rx}{2} \left[ \frac{1}{2} +  \ln\left(\Hpp\over \Hra\right)  \right].
\eea
This indicates that the total field excursion is $\Delta \phi /f = {\cal O}(r_\xi)$, and it only logarithmically depends on model parameters. 
For the following discussion, we choose dark photon coupling such that $r_\xi = \xi / r_X = {\cal O}(\delta)$, indicating that the particle production is triggered before the reappearance of the backreaction potential, and this field excursion can be directly interpreted as misalignment angle $(\Delta \theta)_{\rm mis} \sim {\cal O}(\delta)$. 
This choice of dark photon parameter corresponds to a hierarchy between the relaxion periodicity $f$ and its coupling to dark photon $f/r_X$, which can be achieved in a technically natural way by clockwork mechanism~\cite{Choi:2014rja,Choi:2015fiu,Kaplan:2015fuy,Giudice:2016yja}.

The relaxion begins to oscillate when the temperature of the universe becomes $T_{\rm os} \simeq \min [ (\Mp m_\phi)^{1/2}, T_{\rm ra}] $. 
If $T_{\rm os} =T_{\rm ra}$, i.e., the relaxion begins to oscillate right after the reappearance of back reaction potential, the relic abundance is estimated as
\bea
\frac{\Omega_\phi}{\Omega_{\rm DM}} \sim 
0.1\,\,
\Lambda_{\rm TeV}^{-\frac{4}{3}}
\Big( \frac{f}{10^{15}\GeV} \Big)^{\frac{10}{3}} 
\Big( \frac{m_\phi}{10^{-5} \eV} \Big)^{\frac{10}{3}},
\eea
where we have chosen $T_{\rm ra} =50 \GeV$. 
If $T_{\rm os} \sim \sqrt{\Mp m_\phi}$, then the relic abundance is estimated as
\bea
\frac{\Omega_\phi}{\Omega_{\rm DM}} \sim 
0.1\,\,
\Lambda_{\rm TeV}^{-\frac{4}{3}} 
\Big( \frac{f}{10^{15}\GeV} \Big)^{\frac{10}{3}}
\Big( \frac{m_\phi}{ 10^{-5} \eV } \Big)^{\frac{11}{6}} .
\eea
Note that for this choice of parameters, $T_{\rm os} \simeq 50\GeV$ and $g_\star(T_{\rm os}) \simeq 97$. 
For both estimation, we have assumed that $(\Delta \theta)_{\rm mis} \simeq 2 \delta$.

We note that, as we consider the coherently oscillating relaxion DM, its mass should be less than $\sim 10 \eV$ in order to be described by classical field~(see for instance~\cite{Kolb:1990vq}).
For this range of relaxion mass, the possible decay channels are into two photons and two dark photons, 
$\Gamma = \Gamma_{\gamma\gamma} + \Gamma_{XX}$.
The decay rate to two dark photon is 
\bea
\!\!\!
\Gamma_{XX} \! &=&\! \frac{r_X^2}{64\pi} \frac{m_\phi^3}{f^2}
\sim \frac{\Lambda_{\rm TeV}^{4/3}}{10^2 \, {\rm Gyr}}
\Big( \frac{10^{11} \GeV}{f} \Big)^{\frac{10}{3}} 
\Big( \frac{m_\phi}{\rm eV} \Big)^{\frac{5}{3}}\!,
\eea
where we have chosen $r_X = \xi / \delta$ and assumed $\xi = 25$. 
As the mixing angle between the relaxion and Higgs can be at most $\sin\theta_{h\phi} \sim v/f$, the partial width to diphoton is subdominant relative to its decay into dark diphoton~\cite{Flacke:2016szy,Frugiuele:2018coc}.
Since the decay of dark matter into relativistic particles affects the spectrum of cosmic microwave background at low-$\ell$ multipoles, the lifetime is constrained as $\Gamma^{-1} > 160$ Gyr~\cite{Audren:2014bca}, only mildly constraining our model's parameter space.

We comment on the possibility of parametric resonance during the oscillation phase of relaxion DM. 
We have assumed that the relaxion oscillates around its local minimum after $T_{\rm os}$, while ignoring particle production during this period. 
For $r_X (\Delta \theta)_{\rm sep} \gtrsim {\cal O}(10)$, it has been demonstrated that  exponential production of dark photons from parametric resonance would lead to a suppression in the resulting DM abundance~\cite{Agrawal:2017eqm, Kitajima:2017peg}. 
This suppression factor is at most $\sim 10^{-2}$ for $r_X (\Delta \theta)_{\rm sep} \sim {\cal O}(10^2)$ so that the relaxion could still be a viable dark matter candidate~\cite{Kitajima:2017peg}.
Nevertheless, we consider $r_X \sim {\cal O}(\xi /\delta)$ so that $r_X (\Delta \theta)_{\rm sep} \simeq \xi$, and thus, the parametric resonance during the coherent oscillation phase can be ignored. \\ 

{\bf Other constraints.}
In the minimal relaxion model with sub-eV relaxion mass, additional  constraints on our scenario are from long range forces, and observational data on the history of astrophysical bodies, such as red giants, and stars on horizontal branch, which has been investigated thoroughly in~\cite{Flacke:2016szy,Choi:2016luu}.
Among them, fifth force experiments are the one that significantly restricts the parameter space in dark photon scenario. 

Another restriction on parameter space may arise from overproduction of dark photon.
In the presence of relaxion-dark photon coupling, the evolution of the relaxion after reheating continuously produces dark photons, and their energy density at $\Hra$ is estimated to be
$\rho_X(\Tra) \sim \rx \mub^4\,$~\cite{Choi:2016kke}.
As we have assumed that the universe is dominated by radiation, we require this energy density to be smaller than the radiation energy density at the time of trapping, $\rho_X(\Tra) \lesssim 3\Mp^2 \Hra^2 \sim \Tra^4$, otherwise the relaxion dominates the total energy density of the universe. 
This condition is also required because the dark photon energy density increases $\Delta N_{\rm eff}$ such that the relaxion DM model becomes incompatible with successful big bang nucleosynthesis (BBN) and the observation of cosmic microwave background (CMB). 
For a given reappearance temperature, this requirement can be translated into
\bea
r_\xi \lesssim g_\star(T_{\rm ra})\Lambda_{\rm TeV}^{-\frac{2}{3}} 
\Big( \frac{T_{\rm ra}}{10\GeV} \Big)^4 
\Big( \frac{\rm GeV^2}{m_\phi f } \Big)^{4/3}.
\eea
We are interested in $r_\xi = \xi / r_X \simeq {\cal O}(\delta)$.
This consideration does not constrain the region of parameter space where the relaxion DM can be realized for $T_{\rm ra} \gtrsim 10 \GeV$.

\begin{figure*}
\centering
\hspace*{-0.8cm}
\includegraphics[scale=0.23]{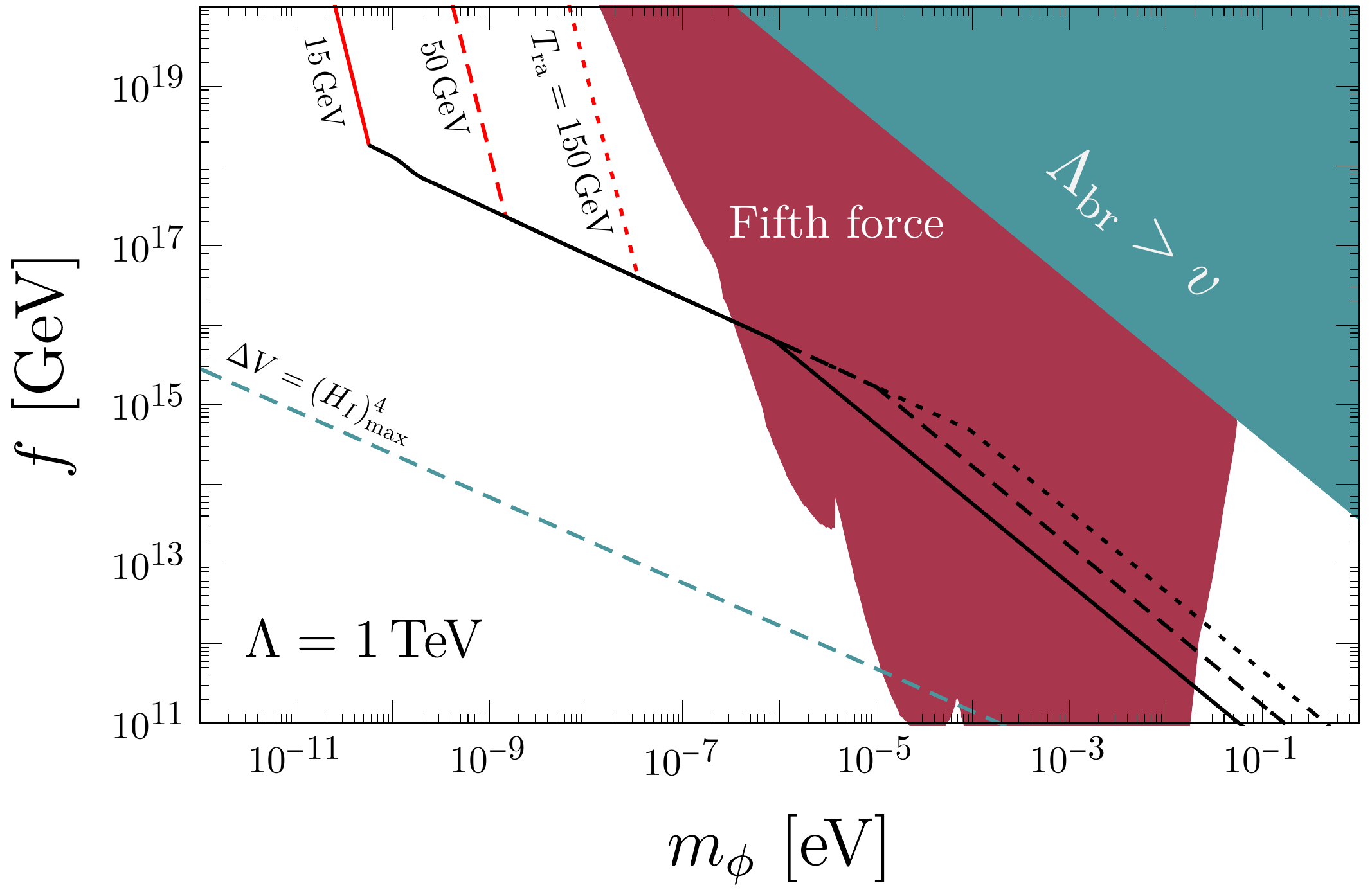}
\quad
\includegraphics[scale=0.234]{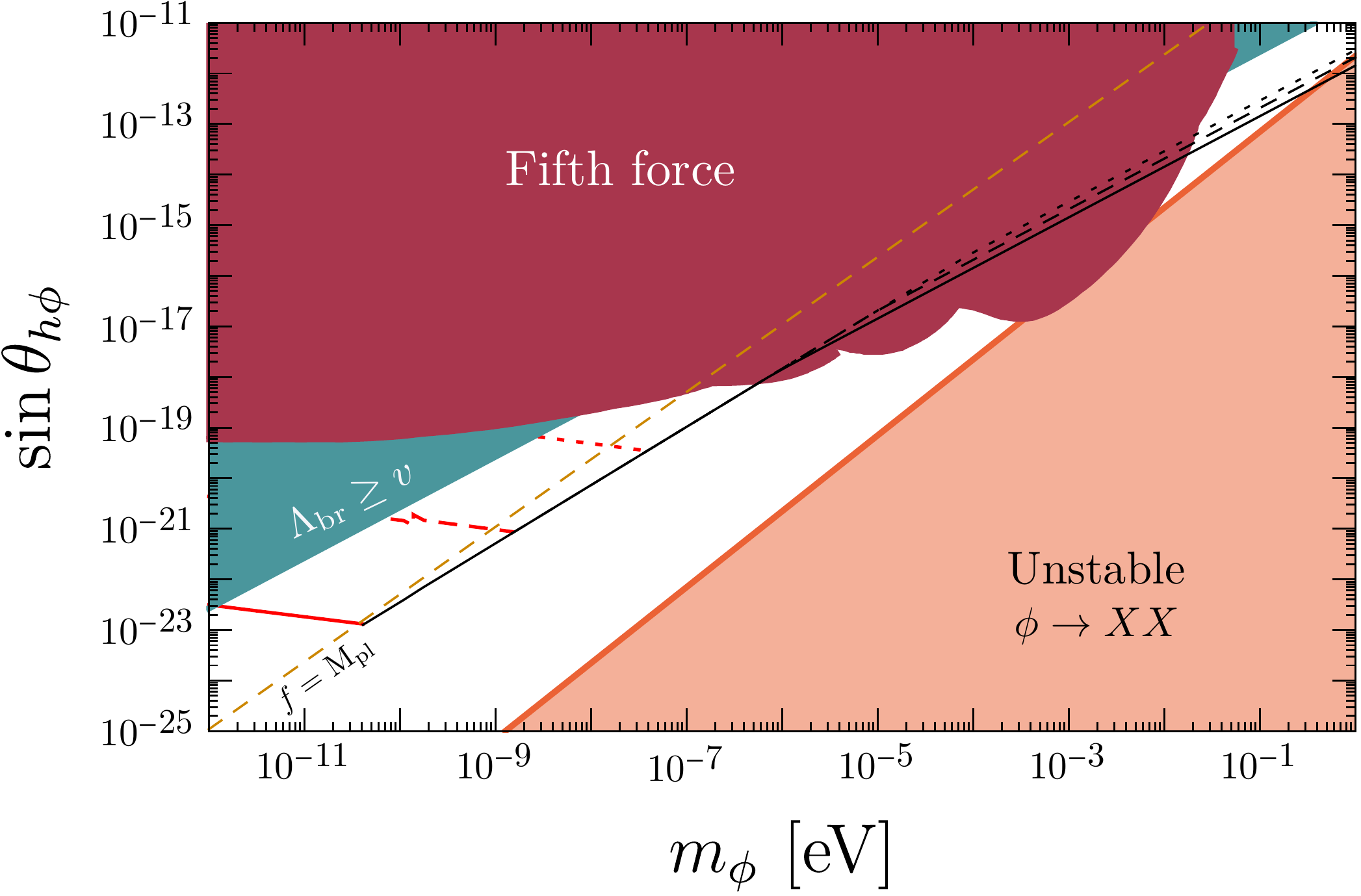}
\caption{
A parameter space for coherent relaxion dark matter in the plane of relaxion mass and decay constant (left), and in the plane of relaxion mass and mixing angle with the Higgs (right).
The cutoff is chosen as $\Lambda = 1\TeV$ for both figures.
The red lines describe regions consistent with the observed DM relic density without dark photon, while the black lines are with dark photon.
We have chosen $\Tra= 150\GeV$ (dotted), $50 \GeV$ (dashed), and $15 \GeV$ (solid). 
The red shaded region is excluded by experiments testing long-range forces~\cite{Hoskins:1985tn,Kapner:2006si,Schlamminger:2007ht} following the procedure described in~\cite{Flacke:2016szy}, and the blue shaded region corresponds to $\mub \gtrsim v$.
The blue dashed line in the left panel corresponds to $\delta^3 \Lambda_{\rm br}^4 = (H_I)_{\rm max}^4 = (\Lambda_{\rm br}^4/f)^{4/3}$ above which the fourth root of the potential barrier at the first local minimum is already larger than the maximum inflationary Hubble scale, $(H_I)_{\rm max}$.
In the orange shaded region (right), the relaxion decays into two dark photons, leaving observable signatures in CMB spectrum.
Above the orange dashed line in the right panel, the relaxion decay constant takes super-Planckian value. 
}
\label{fig:mvsf_rx1}
\end{figure*}

In Fig.~\ref{fig:mvsf_rx1}, we present the region of parameter space where the relaxion could be coherent dark matter and also the relevant constraints on our scenario.
The black and red lines represent $\Omega_\phi= \Omega_{\rm DM}$ for a given choice of relaxion parameters and $\Tra$.
It shows two characteristic behaviors depending on the mass of relaxion.
For the light relaxion mass, $m_\phi \lesssim 3\delta \Hra$, the Hubble friction itself is enough to trap the relaxion at the reappearance time. 
This minimal scenario corresponds to the red lines in both figures. 
For relatively heavier relaxion mass, $m_\phi \gtrsim 3 \delta \Hra$, the Hubble friction is not sufficient to trap the relaxion.
However, in the presence of dark photon production, the evolution of relaxion could still be controlled in a way that it evolves less than $(\Delta \theta)_{\rm sep}$ from the reheating to the reappearance. 
After being trapped, the relaxion starts to oscillate around its minimum at $\Tos$.
Some region of parameter space  for dark photon scenario is already incompatible with various considerations, for instance, fifth force experiments (red shaded), $\Lambda_{\rm br} > v$ (blue shaded), and relaxion decays into two dark photons (orange shaded in the right panel). 
Note also that the blue dashed line in the left panel corresponds to $\Delta V = \delta^3 \Lambda_{\rm r}^4 = (H_I)_{\rm max}^4$, from which we see that, for all of available parameter space of relaxion dark matter, the potential height at the first local minimum is already larger than the maximum $H_I^4$. 
\\

{\bf Discussion.}\label{sec:conclusion}
We have shown how relaxion models with large reheating temperatures can reproduce the observed dark matter relic abundance.
The relaxion behaves as a classical field similar to axion models but does not require any specific value of misalignment angle as an initial condition. 
On the one hand, the relaxion is an axion-like particle with its mass protected by an approximate shift symmetry, while, on the other hand, the relaxion mixes with the Higgs boson and behaves as a classical coherent scalar DM despite the fact that it has nothing to do with the dilaton or scale-invariance symmetry.
The physical relaxion is not a CP eigenstate, and this allows relaxion to have both scalar-coupling and pseudoscalar-coupling to SM particles.
Due to axion-like coupling to SM particles, axion DM searches such as CASPEr, GNOME etc. can be applied to our relaxion DM scenario (see {\it e.g.}~\cite{Graham:2015ouw} and references therein). These experiments are trying to probe various spin dependent effects induced by axions and/or axion like particles.
More interestingly, due to the mixing with the Higgs, fundamental parameters in SM, such as the mass of SM fermions and the fine structure constant, could oscillate because of coherent oscillation of relaxion DM (effect that is far larger than the one resulting from the coupling in charge of the scanning of the Higgs mass~\cite{Graham:2015cka}). 
A change of fundamental constants induced by coherent dark matter field is being actively investigated with precision measurements, such as atomic  clocks~\cite{Arvanitaki:2014faa,Hees:2018fpg,Safronova:2017xyt}, as the energy level of atomic system changes according to the dark matter oscillation.
Although it is challenging to probe oscillations of fundamental constants induced by local DM at these frequencies, a few recently proposed techniques~\cite{Aharony:2019iad,Antypas:2019qji} might be able to probe a certain part of parameter space in special cases where relaxion DM forms compact objects, gravitationally bounded by the solar system~\cite{Banerjee:2019epw}. 

We finally note that Ref.~\cite{Fonseca:2018kqf}, which discusses the possibility of the relaxion being a particle dark matter with low reheating temperature, $T_{\rm rh} \sim 30\MeV$ which is very close to the BBN temperature, appeared, while our paper was in its last phase of preparation.
\vspace*{.51cm}

\acknowledgements
We are grateful for useful discussions and comments on the manuscript from Nayara Fonseca and Oleksii Matsedonskyi. 
We also thank the Galileo Galilei Institute for Theoretical Physics for the hospitality and the INFN for partial support during the completion of this work.
This work was supported by a grant from the Simons Foundation (341344, LA).
The work of GP is supported by grants from the BSF, ERC, ISF the Minerva Foundation, and the Segre Research Award. 

\appendix
\section{Equations of motion and asymptotic behavior of relaxion}\label{sec:app}
The equations of motion for relaxion and dark photon are given as
\bea
0 &=& \ddot{\phi} + 3 H \dot{\phi} + \frac{\partial V(v,\phi)}{\partial \phi}  + \frac{r_X}{4f a^4} \langle X_{\mu\nu} \widetilde{X}^{\mu\nu} \rangle,
\\
0 &=& X_\pm'' + (k^2 \mp  r_X k \theta' ) X_\pm,
\label{dp_eom}
\eea
where the prime and overdot denote a derivative with respect to the conformal time and the physical time, respectively, and $\theta \equiv \phi /f $.
The metric is given as
\bea
ds^2 = dt^2 - a^2(t) \delta_{ij} dx^i dx^j. 
\eea
To investigate how the particle production affects the relaxion evolution, it is more convenient to write the source term in the relaxion equation of motion in Fourier space,
\bea
\frac{1}{4 a^4} \langle X_{\mu\nu} \widetilde{X}^{\mu\nu} \rangle =  \frac{1}{a^4} \int \frac{d^3k}{(2\pi)^3} {k \over 2} \sum_{\lambda=\pm} \lambda \frac{d}{d\tau} |X_\lambda|^2 .
\eea
Since the relaxion velocity is $\theta' > 0 $ in our convention, only $\lambda = +$ helicity is exponentially produced, while $\lambda = -$ helicity state remains almost vacuum fluctuation.

The relaxion evolution before the particle production is dominantly governed by the slope of the relaxion potential. 
At the very beginning of relaxion evolution, its solution in radiation dominated universe is approximated as 
\bea
\dot{\phi}(t) &=& \frac{2}{5} g \Lambda^3 t \bigg[ 1 - \Big(\frac{t_{\rm rh}}{t}\Big)^{5/2} \bigg],
\label{rel_vel_no}
\\
|\Delta \phi(t)| &=& \frac{1}{5} g \Lambda^3 t^2 \left[ 1 - 5 \Big(\frac{t_{\rm rh}}{t} \Big)^2 + 4 \Big( \frac{t_{\rm rh}}{t} \Big)^{5/2} \right],
\label{rel_ex_no}
\eea
where $t_{\rm rh}$ is the physical time at the reheating.
Using this approximate solution, we can estimate the time scale that the particle production becomes important.
For this purpose, we use WKB approximation to solve the equation of motion for dark photon, and find  
\bea
X_+ (k,\tau) \approx \frac{e^{ \int^\tau d\tau' \, \Omega_k (\tau') }}{\sqrt{2\Omega_k(\tau)}}  \equiv  \frac{e^{g_k(\tau)}}{\sqrt{2\Omega_k(\tau)}},
\eea
where the frequency is defined as $\Omega_k^2(\tau) = r_X k \theta' -k^2$, and $g_k(\tau) \equiv \int^\tau d\tau' \, \Omega_k(\tau')$. 
This approximation is valid only when $|\Omega_k' / \Omega_k^2| \ll 1$, which is translated into
\bea
\frac{1}{4r_X} \left| \frac{\theta''^2}{\theta'^4} \right|< k/ | \theta' | < r_X .
\eea
Substituting this solution to the source term, we find
\bea
\frac{1}{4a^4} \langle X_{\mu\nu} \widetilde{X}^{\mu\nu} \rangle 
\approx \frac{1}{4\pi^2 a^4} \! \int dk \, k^3 e^{2 g_k(\tau)}
\sim \frac{k_*^4}{4\pi^2 a^4}  e^{2 g_{k_*}(\tau)}.
\nonumber
\eea
For the last expression, we take a specific wavenumber for the estimation, $k_* = r_X | \theta'(\tau)|$, which becomes stable at $\tau$.
With this wavenumber, we estimate the source term as
\bea
 \frac{1}{4a^4} \langle X_{\mu\nu} \widetilde{X}^{\mu\nu} \rangle 
\!\sim\! \frac{r_X^4 |\dot{\theta}(t)|^4}{4\pi^2}  \exp \bigg[ \frac{4r_X}{5} \frac{\dot{\theta}(t)}{H} \bigg] .
\label{before_pp}
\eea
When this term becomes comparable to the other terms in the equation of motion, for instance, $\partial V(v,\phi) / \partial \phi$, the dark photon begins to affect the relaxion evolution.
Equating the source term with the slope of relaxion potential, we estimate the Hubble scale at the particle production as
\bea
H_{\rm pp} = m_0 \sqrt{\frac{r_X}{5\xi}},
\eea
where $\xi$ is given as a solution of
\bea
\xi = \frac{5}{2} \ln \left( \frac{10\pi}{r_X^{3/2} \xi} \frac{f}{m_0} \right) \sim{\cal O}(10 \,\,\textrm{--} \,\,10^2).
\eea

The relaxion evolution after this time scale is difficult to estimate as the equation becomes integro-differential equation.
Still, we know that the relaxion field velocity must decrease with time.
To see this, we consider the constant relaxion kinetic energy.
In this case, the dark photon field with a constant internal of wavenumber, $0\leq k \leq r_X |\theta'|$, experiences tachyonic instability at a constant rate.
This indicates that these wavenumbers of dark photon keep being exponentially produced, leading to a exponentially growing source.
Therefore, the source term cannot asymptotes to the slope of relaxion potential with a constant relaxion field velocity.

The relaxion field velocity must be a decreasing function in time.
Since we do not know the form of the asymptotic solution, we introduce an ansatz for relaxion evolution, and ask which ansatz satisfies the equation of motion asymptotically.
We introduce
\bea
\theta'(\tau) = \theta'(\tau_{\rm pp}) \left( \frac{\tau_{\rm pp}}{\tau} \right)^n,
\eea
where the subscript indicates a value computed at the time scale of particle production. 
It is interesting to observe that $n=1$ allows scale invariant production of dark photon.
From Eq.~\eqref{dp_eom}, at $k\tau =$ constant surface, the dark photon field is enhanced exactly the same amount relative to its vacuum fluctuation regardless of wavenumber. 
However, $n=1$ cannot be the asymptotic solution because $\langle X \widetilde{X}\rangle$ contributes to relaxion equation of motion with $a^{-4}$, and also because the exponentially produced wavenumber redshifts such that the integral over tachyonic wavenumber also scales as $a^{-4}$.
In other words, despite that the dark photon field, $\sqrt{2k}X(k,\tau)$, is enhanced exactly by the same amount, its contribution to relaxion equation of motion redshifts as $\propto 1/a^8$, while $\partial V(v,\phi) / \partial \phi$ remains as constant.
The relaxion evolution that allows the dark photon source term to asymptote to the slope of the potential would be the one with $n<1$.

To estimate this more carefully, we use the WKB solution again in addition to saddle point approximation, and find
\bea
\!\!\!\!\!\!\!\!\!\!\!\! \frac{1}{4 a^4} \langle X_{\mu\nu} \widetilde{X}^{\mu\nu} \rangle 
\simeq  \frac{e^{ 2  g_{\tilde{k}} (\tau)}}{8\pi^{3/2}}  \bigg( {\tilde{k} \over a} \bigg)^4 
\left( - \tilde{k}^2 \frac{ \partial^2 g_{k}}{ \partial k^2} \bigg|_{\tilde k} \right)^{-\frac{1}{2}} \!\!.
\label{expanded_source}
\eea
The saddle point $\tilde k$ is obtained as a solution of $\partial g_{k}(\tau) / \partial k|_{k=\tilde{k}} = 0$, and is generally a function of conformal time.
To find the asymptotic solution, it is crucial to know how $g_{\tilde k}(\tau)$ scales because the source is exponentially sensitive to it.
By definition, the saddle point satisfies
\bea
\int^\tau d\tau' \, \frac{r_X \theta' (\tau') - 2 \tilde{k}}{2\sqrt{r_X \theta' (\tau') {\tilde k} - {\tilde k}^2}} = 0,
\label{saddle}
\eea
for any conformal time.
If we shift $\tau \to \lambda \tau$ with a scaling parameter $\lambda$, the saddle point should scales as
\bea
\tilde{k} \to \tilde{k}_\lambda = \lambda^{-n} \tilde{k},
\eea
in order to satisfy Eq.~\eqref{saddle}.
This immediately leads to
\bea
\!\!\!\!\!\!\!\!\! g_{\tilde k}(\tau) \to \lambda^{1-n} g_{\tilde k}(\tau),
\quad
\frac{\partial^2 g_k}{\partial k^2}\bigg|_{\tilde{k}} \!\!\!\!\! \to \lambda^{n+1}\frac{\partial^2 g_{ k}}{\partial k^2}\bigg|_{\tilde{k}} \, .
\eea
Substituting this scaling behavior into the source term, Eq.~\eqref{expanded_source}, we notice that the source is a decreasing polynomial of the scaling parameter for $n\geq1$, while it exponentially grows for $n \ll 1$.
Thus, it is $0<1-n \ll1$ that allows the source term to asymptotes to the slope of the relaxion potential. 

In this respect, we write $1-n \equiv \epsilon \ll 1$.
We find
\bea
\frac{a^{-4}\langle X_{\mu\nu} \widetilde{X}^{\mu\nu} \rangle (\tau)}{a_{\rm pp}^{-4}\langle X_{\mu\nu} \widetilde{X}^{\mu\nu} \rangle(\tau_{\rm pp})}
\approx  \left( \frac{\tau}{\tau_{\rm pp}} \right)^{-8 +2 \epsilon g_{\tilde k}(\tau_{\rm pp})}.
\eea
From this, we find
\bea
\epsilon \approx \frac{4}{g_{\tilde k}(\tau_{\rm pp})}  \sim \frac{5}{\xi} < 1 ,
\eea
in order for $a^{-4} \langle X\widetilde{X} \rangle$ to asymptote to the slope of relaxion potential.
Here, we have estimated the exponent $g_{\tilde k}(\tau_{\rm pp})$ from Eq.~\eqref{before_pp}.
Using this result, we find the asymptotic relaxion evolution as
\bea
\!\!\theta' &\simeq& \theta'_{\rm pp} \left( \frac{\tau_{\rm pp}}{\tau} \right)^{1-\epsilon} = r_\xi (aH) \left( \frac{\tau}{\tau_{\rm pp}} \right)^\epsilon,
\\
\!\!\dot{\theta} &\simeq&  r_\xi H \left( \frac{t}{t_{\rm pp}} \right)^{\epsilon/2} 
\simeq r_\xi H \left[ 1 + \frac{\epsilon}{2} \ln\bigg(\frac{t}{t_{\rm pp}}\bigg) \right],
\eea
where $r_\xi \equiv \xi / r_X$. 
The relaxion evolution after the particle production scales as $\dot{\theta} \propto H$ in addition to small logarithmic time dependence, which we ignored in the main text.

\section{Relaxion potential around local minima}\label{sec:app2}
We investigate below the shape of the relaxion potential near local minima.
It is important to correctly describe the potential near local minima because the distance between local extremum determines the maximum misalignment angle, the potential height determines the maximum mass density that the relaxion potential can store, and finally the relaxion mass determines the time of oscillation.
One of the technical difficulties in determining the relaxion quantities near local minimum is that the relaxion stops classical evolution during the inflation at the point where the rolling potential finely balances the backreaction potential.
The overall shape of the potential near local minimum is substantially distorted from $V_{\rm br}$ by the rolling potential, and as a consequence, the relaxion quantities, such as its mass and potential height, could be several orders of magnitude different from those naively computed from $V_{\rm br}$. 

We first discuss how the first local minimum appears in the relaxion scenario, and the relaxion qunatities at this point.
Before the relaxion finds the electroweak scale Higgs mass, the slope of the relaxion potential is always $V'(\phi) <0$ in our convention.
Because of this potential slope, the relaxion evolves, while the Higgs mass and the backreaction potential gradually grow. 
Eventually, at some point in the field space, we will be able to find a solution to $V'(\phi) = 0$ for the first time, and the solution to this equation is the first local minimum in the relaxion field space where the relaxion stops its classical evolution.

We want to compute the relaxion mass, the potential height, and the distance between local extremum at this point. 
For more quantitative discussion, we denote $\phi /f = 2 \pi m + \theta$ where $m$ is an integer number and $\theta \in [ 0 , 2\pi]$. 
We also define $m_*$ to be the smallest integer number that $V'(m_*,\theta) =0$ has a solution for $\theta \in [ 0, 2\pi]$.
To compute the relaxion quantities, we first observe that $\max V'(\theta)|_{m=m_*} ={\cal O}( \delta^2)$.
This is quite natural because, for each $\Delta m=1$, the Higgs mass changes by $\Delta v^2 /v^2 = (\mu/\Lambda)^2 = \delta^2$ and so does the backreaction potential, $\Delta V'_{\rm br} / V'_{\rm br} = \delta^2$. 
With this observation, we expand the first derivative of the relaxion potential around its maximum,
\bea
V'(\theta) \approx \max V'(\theta_*) + \frac{V'''(\theta_*)}{2} (\theta - \theta_*)^2 + \cdots.
\label{approx_struc}
\eea
Thus, the distance between local maximum and minimum, which will become maximum misalignment angle, is
\bea
(\Delta \theta)_{\rm sep} = 2 \left( \frac{2 \max V'}{|V'''| } \right)^{1/2} \sim \frac{2 \delta }{\sqrt{\sin\theta_*}}.
\eea
The physical relaxion mass at the local minimum can be obtained by taking another derivative on Eq.~\eqref{approx_struc}, 
\bea
m_\phi^2 \simeq \frac{1}{2} (\Delta \theta)_{\rm sep} V'''(\theta_*) \sim \frac{\delta m_0^2}{\sqrt{\sin\theta_*}} ,
\eea
while the potential height between local extremum can be obtained by integrating Eq.~\eqref{approx_struc} over $|\theta - \theta_*| < (\Delta \theta)_{\rm sep}/2$,
\bea
\Delta V \simeq (\Delta \theta)_{\rm sep} \max V'(\theta_*) \sim \frac{\delta^3 \Lambda_{\rm br}^4}{\sqrt{\sin\theta_*}}.
\eea
The field value $\theta_*$ is the value at which $V''=0$, or in other words, the slope of the potential $V'$ is maximized within $\theta \in [ 0 , 2\pi]$.
Since the maximum of $V'$ coincides with the maximum of $V'_{\rm br}$, we see that $\theta_* \simeq \pi/2$, and thus, $\sin\theta_* \sim 1$ [Here, we assume $\mu <v$ such that the Higgs VEV can be approximated as a constant. If $v\lesssim \mu \lesssim 4\pi v$, then the Higgs VEV itself has a nonnegligible dependence on the relaxion field, $\langle |H|^2\rangle = ( v^2 + \mu^2 \cos\theta)/2$. As a consequence, the slope of the potential is maximized at $\theta_* \simeq \pi/4$ instead of $\pi/2$].
It is straightforward to generalize these expression for $n$-th local minimum by noting that $\max V' = {\cal O}(n \delta^2)$. 
Thus, at $n$-th minimum, we can still use all of the above expressions by replacing $\delta \to \delta_n = \sqrt{n}\delta$.

\bibliographystyle{h-physrev5}
\bibliography{ref}
\end{document}